\documentclass[journal]{IEEEtran}
\usepackage{algorithm,amsbsy,amsmath,amssymb,epsfig,bbm,mathrsfs,multirow,amsthm}
\usepackage{array,multirow,graphicx}
\usepackage{graphicx}
\usepackage{graphics}
\usepackage{subfigure}
\usepackage{epstopdf}
\usepackage{color}
\usepackage{bbm}
\usepackage{comment}
\usepackage{cite}
\usepackage{algpseudocode}

\renewcommand{\baselinestretch}{1.0}

\newcommand{\beq}{\begin{equation}}
\newcommand{\eeq}{\end{equation}}
\newcommand{\bitm}{\begin{itemize}}
\newcommand{\ba}{\begin{array}}
\newcommand{\ea}{\end{array}}
\newcommand{\eitm}{\end{itemize}}
\newcommand{\beqn}{\begin{eqnarray}}
\newcommand{\eeqn}{\end{eqnarray}}
\newcommand{\beqno}{\begin{eqnarray*}}
\newcommand{\eeqno}{\end{eqnarray*}}
\newcommand{\bma}{\begin{displaymath}}
\newcommand{\ema}{\end{displaymath}}
\newcommand{\bnu}{\begin{enumerate}}
\newcommand{\enu}{\end{enumerate}}
\newcommand{\bce}{\begin{center}}
\newcommand{\ece}{\end{center}}
\newcommand{\btb}{\begin{tabular}}
\newcommand{\etb}{\end{tabular}}

\begin{document}
\title{\huge Efficient Training Management for Mobile Crowd-Machine Learning: A Deep Reinforcement Learning Approach}

\author{ 
\IEEEauthorblockN{Tran The Anh,
Nguyen Cong Luong,
Dusit Niyato~\IEEEmembership{Fellow,~IEEE},
Dong In Kim~\IEEEmembership{Fellow,~IEEE}
and
Li-Chun Wang~\IEEEmembership{Fellow,~IEEE}}

\thanks{T. T. Anh, N. C.~Luong and D. Niyato are with the School of Computer Science and Engineering, Nanyang Technological University, Singapore. Emails: anhtt17895@gmail.com, clnguyen@ntu.edu.sg, dniyato@ntu.edu.sg.}
\thanks{D.~I.~Kim is with School of Information and Communication Engineering, Sungkyunkwan University, Korea. Email: dikim@skku.ac.kr.}
\thanks{L.-C.~Wang is with the Department of Electrical and Computer Engineering, National Chiao Tung University, Taiwan. E-mail: lichun@cc.nctu.edu.tw.}
\vspace{-0.7cm}
}
\maketitle
\begin{abstract}
In this letter, we consider the concept of \textit{Mobile
Crowd-Machine Learning} (MCML) for a federated learning model. The
MCML enables mobile devices in a
mobile network to collaboratively train neural network models
required by a server while keeping data on the mobile devices.
The MCML thus addresses data privacy
issues of traditional machine learning. However, the mobile devices are constrained by energy, CPU,
and wireless bandwidth. Thus, to minimize the
energy consumption, training time and communication cost, the
server needs to determine proper amounts of data and energy
that the mobile devices use for training. However, under
the dynamics and uncertainty of the mobile environment, it is
challenging for the server to determine the optimal decisions on
mobile device resource management. In this letter, we propose
to adopt a deep-Q learning algorithm
that allows the server to learn
and find optimal decisions without any a priori knowledge of
network dynamics. Simulation results show that the proposed algorithm outperforms the static
algorithms in terms of energy consumption and training latency.  \vspace{-0.2cm}
\end{abstract}
\begin{IEEEkeywords}
Mobile crowd, federated learning, deep reinforcement learning.
{\vspace{-0.4cm}}
\end{IEEEkeywords}

\section{Introduction}

Traditional machine learning approaches train neural network models by using data, i.e., training data or dataset, that is centralized in a server or a data center. However, such an approach faces several critical issues related to data privacy, e.g., sensitive data leakage. For example, consider the training of a deep Neural Network (NN) model to predict the word that a user will type when composing a text message on its mobile phone~\cite{goodman2002}. To achieve the high prediction, the server uses a large number of text messages uploaded by the users to train the model. However, the text messages often contain sensitive information, and the users may be reluctant to upload them to the server.


To address the privacy issues, a decentralized machine learning paradigm has been recently proposed that is called \textit{federated learning}~\cite{google2017}. In the federated learning, the mobile devices use their local data to train the model required by the server. The mobile devices then send the model updates, i.e., the model's weight parameters, rather than the data to the server. Federated learning becomes technically feasible due to the fact that mobile devices become more powerful in terms of storage capacity and computation capability. Federated learning can be called \textit{Mobile Crowd-Machine Learning} (MCML) as the learning task is performed by a mobile crowd network. The MCML allows the mobile devices to collaboratively train the shared model while keeping the data on the mobile devices, alleviating the privacy issues. 

However, the MCML has two major limitations. First, the mobile devices have energy and CPU constraints that may reduce the network lifetime and efficiency of training tasks. Second, significant wireless communications often are required for uploading and downloading the model parameters that increases the bandwidth cost and the training latency. To overcome these limitations, the server should decide how much data, energy and CPU resources used by the mobile devices such that the energy consumption, training latency, and bandwidth cost are minimized while meeting requirements of the training tasks. However, it is challenging for the server to determine the optimal decisions since the mobile crowd environment is stochastic in which the energy and CPU states of the mobile devices are uncertain. In this letter, we thus propose to use the Deep Q-Learning (DQL) technique~\cite{van2016deep} that enables the server to find the optimal data and energy management for the mobile devices participating in the MCML through federated learning without any a priori knowledge of network dynamics. We first formulate the stochastic optimization problem for the MCML. Then, we adopt the DQL based on the Double Deep Q-Network (DDQN) to achieve the optimal policy for the server. Simulation results show that the proposed DQL outperforms the non-learning, static algorithms in terms of energy consumption and training latency.  
\vspace{-0.3cm}
\section{System Model}
\vspace{-0.1cm}
\subsection{Federated Learning}
Federated learning~\cite{google2017} is a decentralized learning technique in which a number of mobile devices use their data to train a model required by a server. Federated learning can be performed by multiple iterations. At each iteration, the mobile devices download a so-called \textit{global model}, i.e., the model's weights, from the server. The mobile devices use their local data, energy, and CPU resources to train and improve the global model. The mobile devices then transmit the model updates, i.e., the learned weights, to the server. The server aggregates the model updates and generates a new global model. The mobile devices and the server can repeat the above process at the next iterations until the global model achieves a certain accuracy required by the server. 
\vspace{-0.3cm}
\subsection{System Description}
\begin{figure}[h]
 \centering
\includegraphics[width=7.3cm, height = 6.2cm]{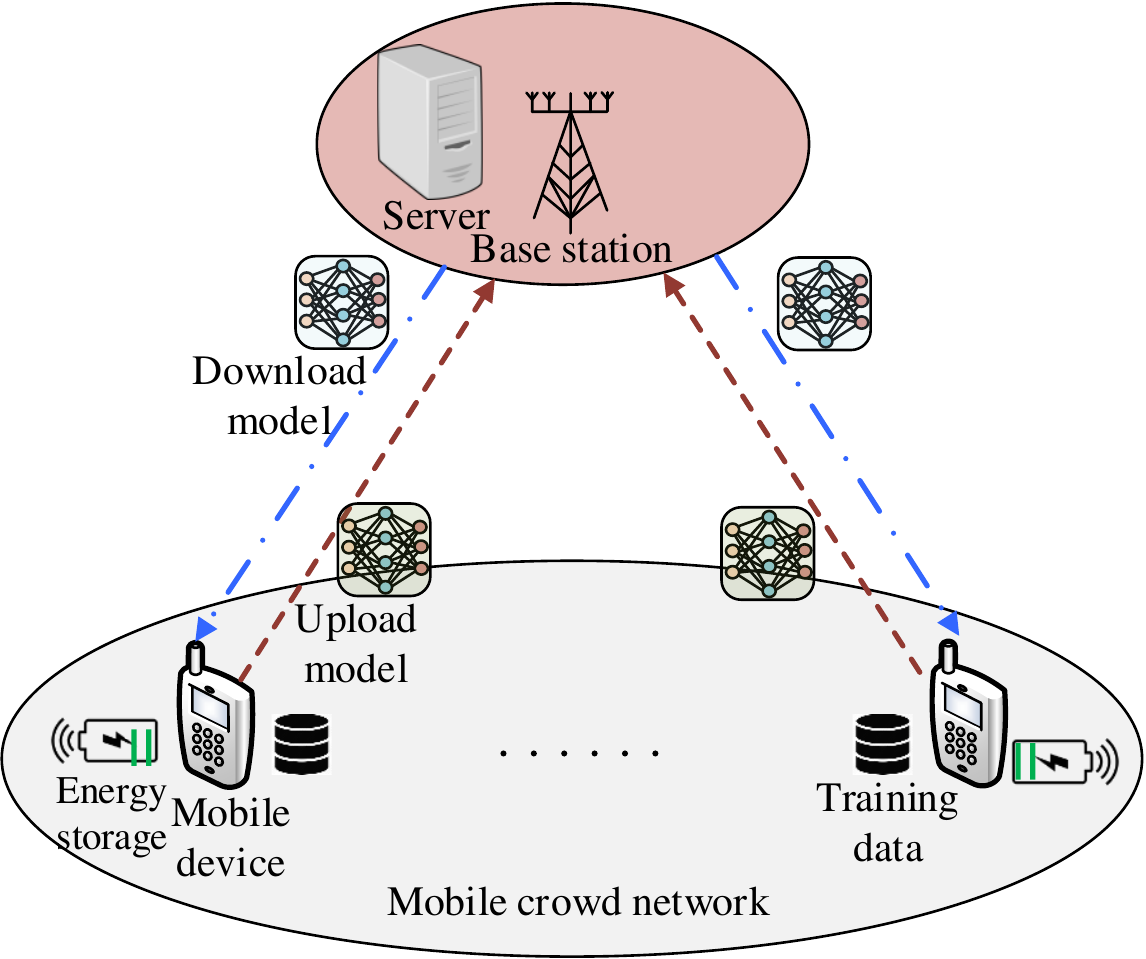}
  \vspace{-0.2cm}
 \caption{Mobile crowd-machine learning system.}
  \label{federated_learning}
  \vspace{-0.5cm}
\end{figure}
In this letter, we consider a federated learning system that uses a mobile crowd network, a self-organized
network with a finite number of mobile devices, for training the model. Such a system is called the MCML as shown in Fig.~\ref{federated_learning}. The system consists of $N$ mobile devices that communicate with the server through a cellular base station. Each mobile device $n$ is equipped with an energy storage. At each iteration, the energy storage has $c_n$ energy units and can be charged with $A_n$ energy units, e.g., from wired or wireless charging facility. Here, one energy unit equals $\delta$ Joule (J). The mobile device has $f_n$ CPU shares, each of which corresponds to $\mu$ CPU cycles. The CPU share is a portion of the mobile device's CPU resource/time used for the training. The mobile devices may have different data quality.~The data quality may be represented by the number of error samples included in the data and is measured, e.g., by using the Alpha test tool \cite{ahonen2017}. At each iteration, the server needs to determine the number of data units $d_n$ and energy units $e_n$ that each mobile device $n$ uses to train the model to minimize the total energy consumption and the training latency while achieving the target accuracy.

\vspace{-0.5cm}
\section{Problem Formulation}
\label{sec:problem_formulation}
The problem of mobile device resource management can be formulated as a stochastic optimization problem defined as $<{\mathcal{S}}, {\mathcal{A}}, {\mathcal{P}}, {\mathcal{R}}>$, where ${\mathcal{S}}$, ${\mathcal{A}}$, ${\mathcal{P}}$, and ${\mathcal{R}}$ are the state space, action space, state transition probability function, and reward function of the server, respectively.
\vspace{-0.5cm}
\subsection{State Space}
The state space of the server is the combination of the state spaces of $N$ mobile devices that is expressed as $\mathcal{S} = \prod_{n=1}^{N}\mathcal{S}_n$, where $\prod$ is the Cartesian product, and $\mathcal{S}_n$ is the state space of mobile device $n$.~$\mathcal{S}_n$ is defined as
\begin{equation}
    \mathcal{S}_n=\big{\{} \left ( f_n, c_n \right );f_n\in \left \{ 0,1,\ldots,F_n \right \}, c_n\in \left \{ 0,1,\ldots,C_n \right \} \big{\}}, \notag
\end{equation}
where $F_n$ is the maximum number of CPU shares of mobile device $n$, and $C_n$ is the capacity of its energy storage.

\vspace{-0.4cm}
\subsection{Action Space}
At each iteration, the server requires $d_n$ data units and $e_n$ energy units from mobile device $n$. Then, the number of CPU cycles required from the mobile device is~\cite{chen2018} $\bar{f}_n = \sqrt{\delta e_n}/\sqrt{\tau \nu d_n}$,
where $\nu$ is the number of CPU cycles required to train one data unit, and $\tau$ is the effective switched capacitance that depends on chip architecture of the mobile device. The action space of the server is defined as follows:
\begin{align}
    \mathcal{A} =&\big{\{}( d_1, e_1,\ldots, d_N, e_N);\\
    &  d_n \leq D_n, e_n \leq c_n, \bar{f}_n \leq \mu f_n, n =(1,\ldots,N) \big{\}},
    \notag
\end{align}
where $D_n$ is the maximum number of data units that mobile device $n$ can use to train the model at the iteration. The constraints $e_n \leq c_n$ and $\bar{f}_n \leq \mu f_n$ ensure that the numbers of energy units and CPU cycles required by the server do not exceed those of the mobile device.

\vspace{-0.3cm}
\subsection{State Transition}
The state transition of the server from the current state $s$ to the next state $s'$ is determined based on the transitions of the energy and CPU states of all the mobile devices. At each iteration, mobile device $n$ consumes $e_n$ energy units for the training and wireless communications, and it may be charged with $A_n$ energy units. Thus, the energy state of mobile device $n$ changes from $c_n$ to $c'_n$ as follows, ${c}'_n=\min \{ c_n - e_n + A_n, C_n \}$, where $A_n$ can follow any distribution. Here, $A_n$ is assumed to follow Poisson distribution with average energy arrival rate $\omega$~\cite{chen2018}, and thus probability that the mobile device charges $k$ energy units is $ P\left (A_n=k \right )= e^{-\omega}\frac{\omega^{k}}{k!}$.

For the CPU state transition, the CPU state of mobile device $n$ changes from $f_n$ to $f'_n$, where $f'_n$ is assumed to follow the uniform distribution $U[0, F_n]$.

\vspace{-0.3cm}
\subsection{Reward Function}
The machine learning model accuracy is proportional to data accumulated from the mobile devices over iterations.~Thus, the reward function should be proportional to the accumulated data and inversely proportional to the energy consumption and training latency. 
\begin{itemize}
\item \textit{Accumulated data ($D$):} Let $\eta _1:\eta_2:\cdots:\eta_N$ denote the data quality ratio among $N$ mobile devices. At each iteration, $D$ can be defined as $ D = \sum_{n=1}^{N} \eta_n d_n/\sum_{n=1}^{N} \eta_n$.
\item \textit{Energy consumption ($E$):} The total energy consumed by $N$ mobile devices for the training at each iteration is $E=\sum_{n=1}^{N}e_n$.
\item \textit{Training latency ($L$):} At the iteration, mobile device $n$ downloads the model from the server, trains the model, and uploads the learned model to the server. Thus, the training latency of the mobile device can be defined as $L_n = \nu d_n/\bar{f}_n + L^{\text{trans}}_n$, where $\nu d_n/\bar{f}_n $ refers to the training time and $L^{\text{trans}}_n$ is the total transmission time for downloading and uploading the model. In general, $L^{\text{trans}}_n$ is proportional to the bandwidth assigned to the mobile device. Thus, the overall training latency is $L = \underset{n}{\max}\left \{ L_n \right \}$.
\end{itemize}

The reward of the server is defined as a function of state $s \in \mathcal{S}$ and action $a \in \mathcal{A}$ as follows:
\begin{equation}
    \mathcal{R}\left ( s,a \right )= \alpha_D \frac{D}{D_{\text{max}}} - \alpha_L \frac{L}{L_{\text{max}}} - \alpha_E \frac{E}{E_{\text{max}}},
\end{equation}
where $\alpha_D, \alpha_L$, and $\alpha_E$ are the scale factors. Here, $D, L$, and $E$ are normalized by their corresponding maximum values. 


To maximize the long-term accumulated reward, the server needs to determine its optimal action $a \in {\mathcal{A}}$ given state $s \in {\mathcal{S}}$. For this, the server determines an optimal policy defined as $\pi^* : {\mathcal{S}} \rightarrow {\mathcal{A}}$.~To find the optimal policy, the Q-learning can be adopted. The Q-learning constructs a look-up table including Q-values of state-action pairs, i.e., $Q(s,a)$. The Q-values are updated based on experiences of the server as follows: 
\begin{equation}
\label{Q_value_update}
Q^{\mathrm{new}}(s,a) =  (1-\lambda) Q(s,a) + \lambda \Big{(} r(s,a) + \gamma \max_{a' \in {\mathcal{A}} } Q( s', a') \Big{)},\notag
\end{equation}
where $\lambda$ is the learning rate, $\gamma$ is the discount factor, and $r(s, a)$ is the reward received.

Based on the look-up table, the server can find its optimal action from any state.~However, the Q-learning suffers from large state and action spaces of the server. Thus, we propose to use the DQL to find the optimal policy for the server. 

\section{Deep Q-Learning Algorithm}

The standard DQL~\cite{mnih2015} uses a single NN in which the input is one of the states of the server and the output includes Q-values of all possible actions of the server. Training the NN enables the server to find optimal actions, i.e., the optimal policy, from its current states. However, the standard DQL with the single NN has the overoptimistic value estimate problem since the DQL uses the same Q-values both to select and to evaluate an action of the server~\cite{van2016deep}. 

To decouple the action selection from the action evaluation, we propose to use the DQL based on DDQN~\cite{van2016deep}. The DDQN consists of two NNs, i.e., the online NN with weights $\boldsymbol{\theta}$ and the target NN with weights $\boldsymbol{\theta^{-}}$.~The online NN updates its weights $\boldsymbol{\theta}$ at each iteration, and the target NN resets its weights $\boldsymbol{\theta^{-}}=\boldsymbol{\theta}$ in every $N^{-}$ iterations. In particular, based on experience $ e=<s,a,r,s'>$ received by the server, the online NN  performs a gradient descent step on the loss function $L(\boldsymbol{\theta})= (y-Q(s,a;\boldsymbol{\theta}))^2$ to update its weights $\boldsymbol{\theta}$. Here, $y$ is the target value that is defined as
\begin{equation}
y=r + \gamma Q\Big{(} s', \arg\max_{a' \in \mathcal{A}} Q(s',a';\boldsymbol{\theta});\boldsymbol{\theta^{-}}\Big{)}.
\label{DQN_y_value_DDQN}
\end{equation}

In (\ref{DQN_y_value_DDQN}), selecting an action is based on weights $\boldsymbol{\theta}$ of the online NN, and evaluating the action uses weights $\boldsymbol{\theta^{-}}$ of the target NN that prevents the overoptimistic problem. 
\begin{algorithm}
\footnotesize 	
\caption{\small DQL algorithm with experience replay~\cite{van2016deep}.}\label{DDQN_Algorithm}
\begin{algorithmic}[1]
\State \textbf{Initialize:} $\boldsymbol{\theta}, \boldsymbol{\theta}^{-}$;
\For{\texttt{episode $i=1$ to $N$}}
\For{\texttt{iteration $t=1$ to $T$}}
\State Select action $a$ according to the $\epsilon$-greedy policy;
\State Execute action $a$ and observe reward $r$ and next state $s'$; 
\State Store experience $e=<s, a, r, s'>$ in $\mathcal{M}$;
\State Select $N_b$ experiences $e_k = <s_k, a_k, r_k, s'_k>$ from $\mathcal{M}$;
 \For{\texttt{$k=1$ to $N_b$}}
  \State Determine $a^{\text{max}}=\arg \max_{a' \in \mathcal{A}} Q(s'_{k}, a'; \boldsymbol{\theta})$;
 \State Calculate $y_k=r_k+ \gamma Q\left (s'_{k}, a^{\text{max}};\boldsymbol{\theta}^{-} \right)$;
 \EndFor 
 \State Define $\overline{L}(\boldsymbol{\theta})=\frac{1}{N_b} \sum_{k=1}^{N_b} \Big{(}y_k -Q(s_k,a_k;\boldsymbol{\theta}) \Big{)} ^{2}$;
\State Perform a gradient descent step on $\overline{L}(\boldsymbol{\theta})$ to update $\boldsymbol{\theta}$;

\State Reset $\boldsymbol{\theta}^{-} = \boldsymbol{\theta}$ in every $N^{-}$ iteration;

\EndFor 
\EndFor
\end{algorithmic}

\end{algorithm}
Algorithm~\ref{DDQN_Algorithm} shows the DQL algorithm for training the DDQN to find the optimal policy for the server. The algorithm consists of the experience phase and the training phase. In the experience phase, the server executes each action $a \in \mathcal{A}$ and receives experience $e$. Here, action $a$ is selected according to the $\epsilon$-greedy policy to balance the exploration and exploitation. To improve the stability of the learning, an experience replay memory $\mathcal{M}$ is used to store the experiences. In the training phase, a mini-batch of $N_b$ experiences $e_k = <s_k, a_k, r_k, s'_k>$ is taken at each iteration to calculate the target values $y_k$ (Lines 11 and 12) and the average loss function $\overline{L}(\boldsymbol{\theta})$ (Line 14). The value of $\overline{L}(\boldsymbol{\theta})$ is used to update weights $\boldsymbol{\theta}$ of the online NN. 

\section{Performance Evaluation}

In this section, we provide the performance evaluation to illustrate the effectiveness of the proposed DQL scheme. The machine learning model required by the server is a Recurrent Neural Network (RNN) for the text prediction application. We perform experiments to measure the accuracy of the prediction based on the amount of local dataset on the mobile devices and energy consumption. The experiment results can be used in the performance evaluation of the mobile device resource management in the MCML framework. For comparison, we introduce the greedy scheme~\cite{devore2017} and the random scheme as baseline schemes. In the greedy scheme, the server executes its action such that the maximum number of data units and CPU shares of each mobile device is taken. In the random scheme, the server selects randomly the number of data units and energy units from each mobile device. Here, we do not use the Q-learning algorithm as it cannot be run in our computation environment due to the high complexity of the problem. Simulation parameters are listed in Table~\ref{table:parameters}. Note that to facilitate the presentation of the results, we consider the small number of mobile devices. The DQL scheme can be applied to the larger number of mobile devices. 

\begin{table}[h]
\vspace{-0.4cm}
\small
\caption{\small Simulation parameters}
\label{table:parameters_CRN}
\footnotesize 	
\centering
\begin{tabular}{lc|lc}
\hline\hline
{\em Parameters} & {\em Value} & Parameters & Value\\ [0.5ex]
\hline
 $N$    & $3$           & $\mu$ &   $0.6$ GHz     \\ 
\hline
 $E_n$    & $3$          & $\delta$ &   $1$      \\ 
\hline
$D_n$  & $3$            &  NN size &   $32\times32\times32$      \\ 
\hline
$\nu$ & $10^{10}$         &  Optimizer&   Adam    \\ 
\hline
$\tau$ & $10^{-28}$  & $\lambda$    & $0.001$     \\ 
\hline
$\gamma$ & 0.9       &  $\epsilon$-greedy &    0.9 $\rightarrow$ 0    \\ 
\hline
\end{tabular}
\label{table:parameters}
\end{table}


\begin{figure}[h]
\vspace{-0.2cm}
 \centering
\includegraphics[width=6.0cm, height = 4.5cm]{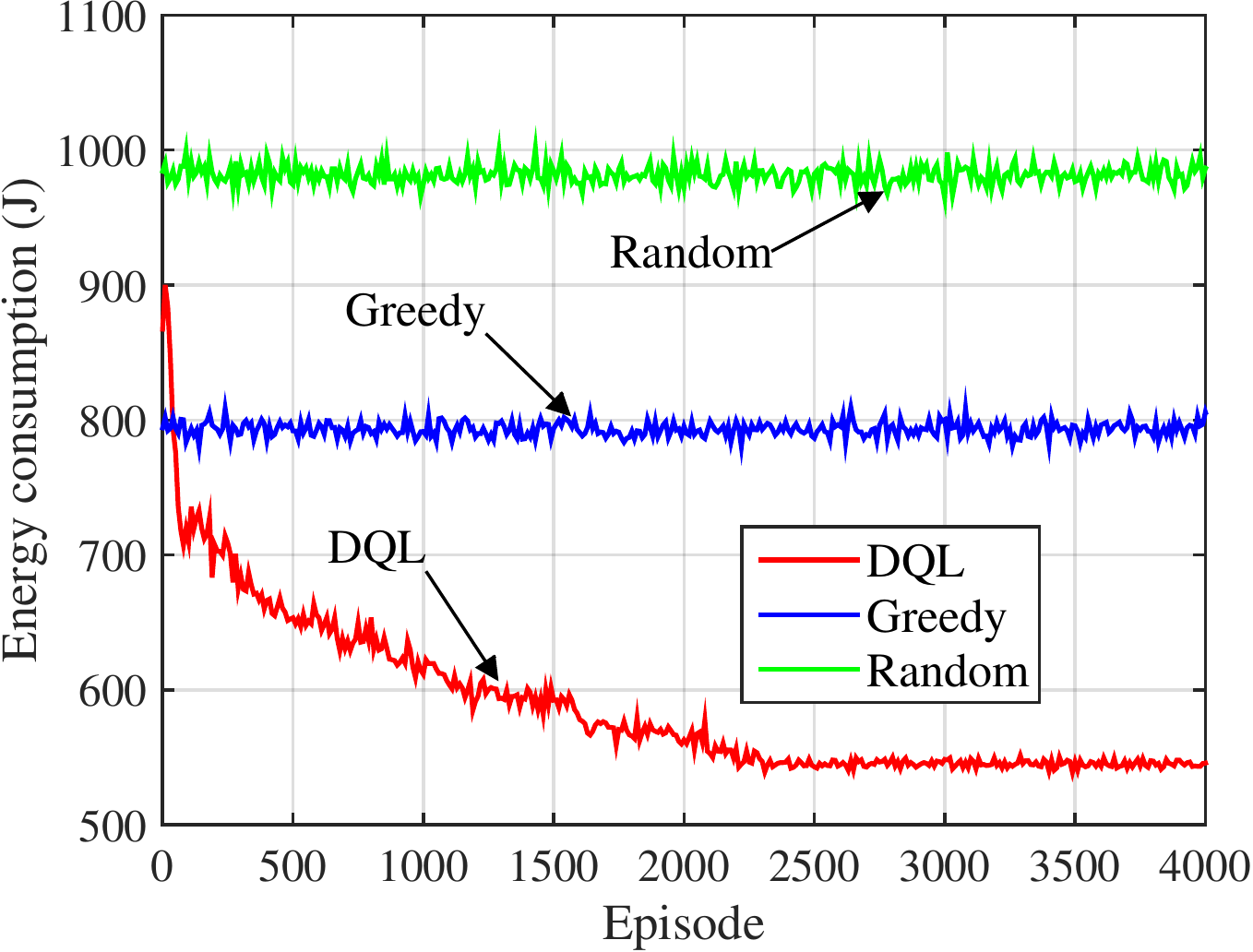}
   \vspace{-0.2cm}
 \caption{Energy consumption comparison.}
  \label{energy_consumption}
     \vspace{-0.6cm}
\end{figure}

Due to the energy constraint of mobile devices, improving energy efficiency is one
of the most important issues in the MCML. Thus, we first consider how the proposed DQL scheme reduces the energy consumption of the network. Fig.~\ref{energy_consumption} shows the total energy consumption of the mobile devices obtained by different schemes. As seen, the DQL scheme converges an energy consumption value that is much lower than those obtained by the baseline schemes. In particular, the DQL scheme reduces the energy consumption around $31\%$ compared with the greedy scheme. The reason can be explained as follows. The greedy scheme always takes the maximum number of data units from the mobile devices, and the mobile devices may consume a large amount of energy for training the data units. With the DQL scheme, the energy cost is taken into account the reward function of the server, and maximizing this function reduces the energy consumption.

\begin{figure}[h]
   \vspace{-0.3cm}
 \centering
\includegraphics[width=5.9cm, height = 4.7cm]{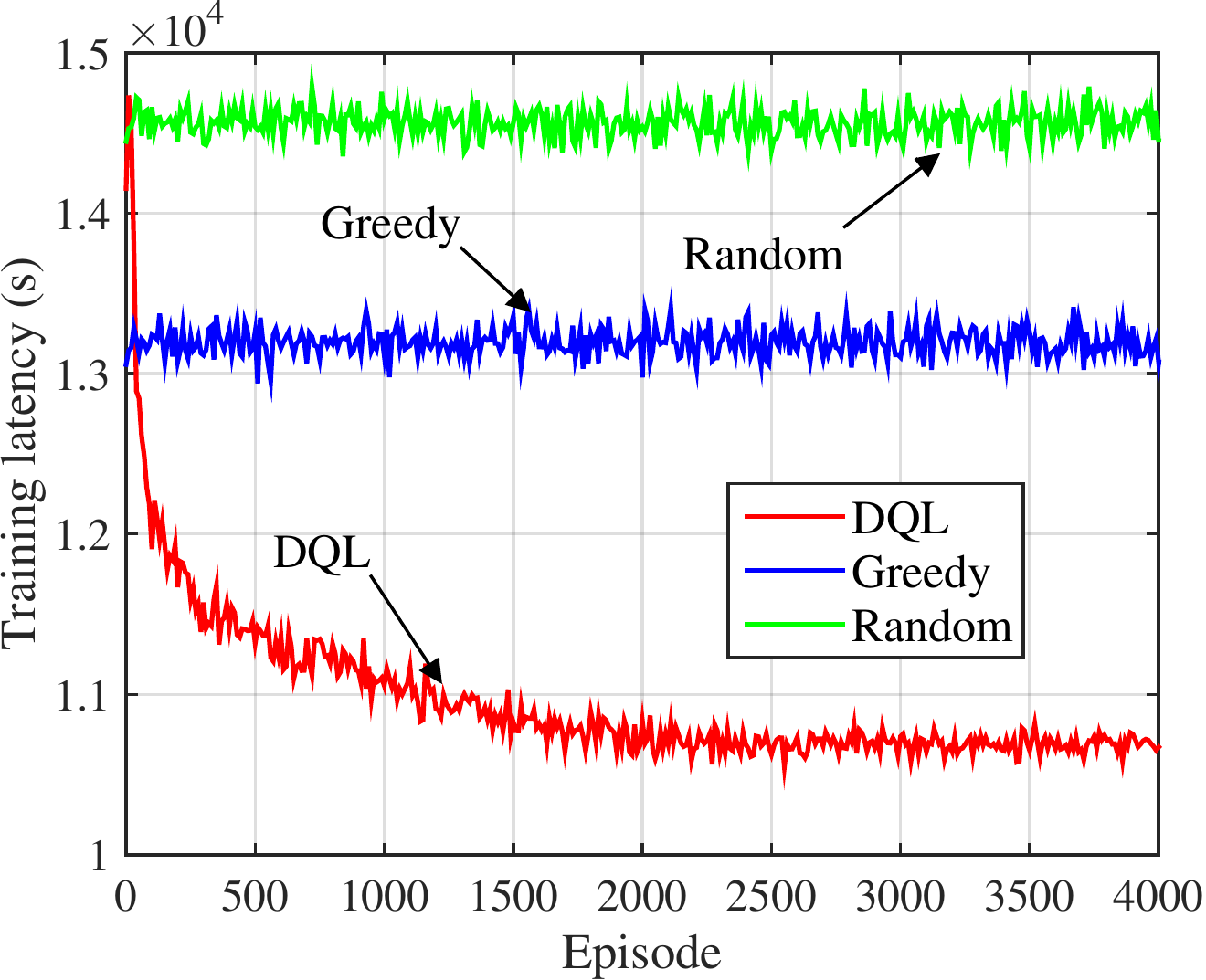}
   \vspace{-0.3cm}
 \caption{Training latency comparison.}
  \label{computational_delay}
     \vspace{-0.2cm}
\end{figure}
Next, we compare the DQL scheme and the baseline schemes in terms of training latency. As shown in Fig.~\ref{computational_delay}, the training latency obtained by the DQL scheme is significantly lower than those obtained by the baseline schemes. The DQL scheme reduces the training latency up to $55\%$ compared with the random scheme. The reason can be explained based on the training latency definition. Accordingly, the training latency depends particularly on the longest training time of the mobile devices at each iteration. As the DQL scheme is used, the server can observe the CPU states of the mobile devices and then decides the proper number of data units such that the mobile devices can finish the training process at the same time. For example, the server takes the low number of data units from the mobile devices that have the small number of CPU shares. As the random scheme is adopted, the server chooses randomly the number of data units from the mobile devices without considering their CPU states. Thus, there is a high probability that the server takes the large number of data units from the mobile devices that have the low number of CPU shares. These mobile devices need much time to finish their training process, resulting in the high training latency.  

\begin{figure}[h]
   \vspace{-0.4cm}
 \centering
\includegraphics[width=5.7cm, height = 4.4cm]{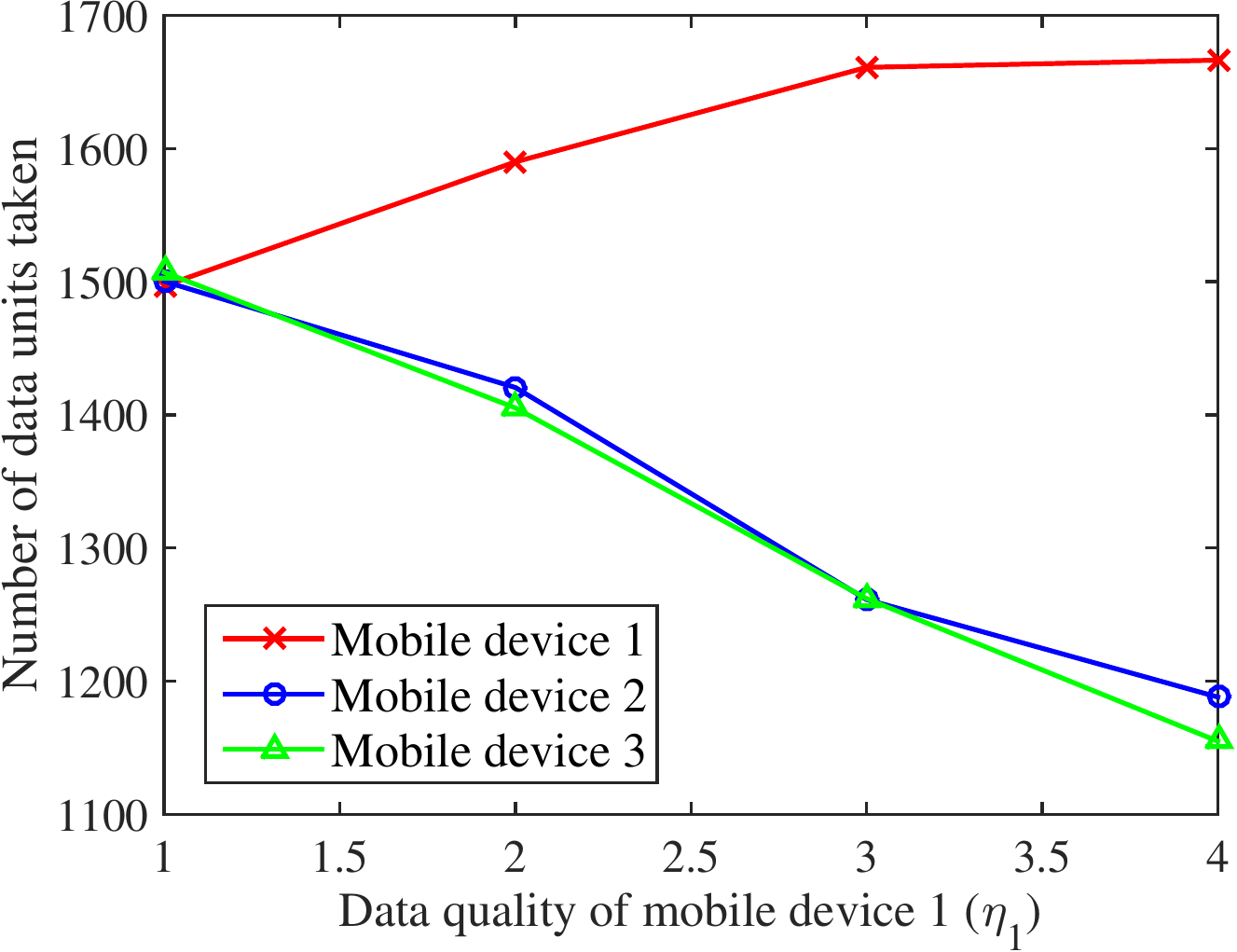}
   \vspace{-0.2cm}
 \caption{The number of data units taken by the server as $\eta_1$ varies.}
  \label{data_quality}
     \vspace{-0.2cm}
\end{figure}
The above results are obtained with the assumption that the mobile devices in the network have the same data quality.~In practice, the data quality among them may be different, and it is essential to see how the server takes the number of data units from the mobile devices. We therefore vary the data quality ratio of mobile device 1 to the other mobile devices. Let $\eta _1:1:1$ denote the data quality ratio among the mobile devices. As shown in Fig.~\ref{data_quality}, as $\eta _1$ increases, the number of data units taken from mobile device 1 increases while those taken from mobile devices 2 and 3 decrease. This simply explains that the server is willing to take more data units from the mobile device with higher data quality such that the server can reach its target accuracy faster. 
\vspace{-0.1cm}
\section{Conclusions}
In this letter, we have presented the DQL scheme for the
optimal data and energy management problem in the mobile crowd-machine learning system. First, we have formulated the stochastic optimization problem for the data and energy management of the server. Then, we have developed a DQL algorithm using DDQN to solve the problem. The simulation results show that the DQL scheme outperforms the non-learning algorithms in terms of energy consumption and training latency.


\end{document}